\documentclass{acm_proc_article-sp}

\newtheorem{theorem}{Theorem}
\newtheorem{lemma}{Lemma}
\newdef{definition}{Definition}
\newdef{note}{Note}

\begin{document}

\conferenceinfo{SAC}{'03 Melbourne, Florida, USA}
\CopyrightYear{2003}

\title{Schedulers for Rule-Based Constraint Programming}

\numberofauthors{2}

\author{
\alignauthor    Krzysztof R.~Apt%
\titlenote{Currently on leave at School of Computing, National University of
 Singapore}\\
        \affaddr{CWI, P.O. Box 94079}\\
        \affaddr{1090 GB Amsterdam, the Netherlands}\\
        \affaddr{and University of Amsterdam, the Netherlands}\\
        \email{K.R.Apt@cwi.nl}
\alignauthor    Sebastian Brand\\
        \affaddr{CWI, P.O. Box 94079}\\
        \affaddr{1090 GB Amsterdam, the Netherlands}\\
        \email{S.Brand@cwi.nl}
}

\maketitle

\newcommand{\friends}   {\ensuremath{\mathit{friends}}}
\newcommand{\obviated}  {\ensuremath{\mathit{obviated}}}
\newcommand{\update}    {\ensuremath{\mathit{update}}}
\newcommand{\holds}     {\ensuremath{\mathit{Holds}}}
\newcommand{\nholds}    {\ensuremath{\neg\mathit{Holds}}}

\newcommand{\GIclassalgo}[1]    {\texttt{#1}}
\newcommand{\GIalgo}            {\GIclassalgo{GI}}
\newcommand{\RGIalgo}           {\GIclassalgo{RGI}}
\newcommand{\Ralgo}             {\GIclassalgo{R}}
\newcommand{\FOalgo}            {\GIclassalgo{F~\&~O}}
\newlength{\figalgogap}\setlength{\figalgogap}{-4.1ex}

\newcommand{\eclipse}   {ECL$^i$PS$^e$}

\newcommand{\ra}{\mbox{$\:\rightarrow\:$}}
\newcommand{\C}[1]{\mbox{$\{{#1}\}$}}
\newcommand{\LL}{\mbox{$\ldots$}}
\newcommand{\po}{\mbox{$\ \sqsubseteq\ $}}
\newcommand{\HB}{\hfill{$\Box$}}
\newcommand{\A}{\mbox{$\ \wedge\ $}}
\newcommand{\fa}{\mbox{$\forall$}}
\newcommand{\ES}{\mbox{$\emptyset$}}
\newcommand{\Or}{\mbox{$\ \vee\ $}}
\newcommand{\sse}{\mbox{$\:\subseteq\:$}}
\newcommand{\te}{\mbox{$\exists$}}
\newcommand{\p}[2]{\langle #1 \ ; \ #2 \rangle}

\def\ts{\textstyle}
\def\smallromani{\renewcommand{\theenumi}{\roman{enumi}}
\renewcommand{\labelenumi}{(\theenumi)}}
\def\smallromaniprime{\renewcommand{\theenumi}{\roman{enumi}\mbox{$'$}}
\renewcommand{\labelenumi}{(\theenumi)}}
\def\smallromanir{\renewcommand{\theenumi}{\roman{enumi}}
\renewcommand{\labelenumi}{\theenumi)}}
\def\smallromanii{\renewcommand{\theenumii}{\roman{enumii}}
\renewcommand{\labelenumii}{(\theenumii)}}
\def\parentarabici{\renewcommand{\theenumi}{\arabic{enumi}}
\renewcommand{\labelenumi}{(\theenumi)}}
\def\parentarabicii{\renewcommand{\theenumii}{\arabic{enumii}}
\renewcommand{\labelenumii}{(\theenumii)}}
\def\eqnum{\renewcommand{\theequation}{\arabic{equation}}
\renewcommand{\labelequation}{(\theequation)}}

\def\parentalphi{\renewcommand{\theenumi}{\alph{enumi}}
\renewcommand{\labelenumi}{(\theenumi)}}

\begin{abstract}
  We study here schedulers for a class of rules that naturally arise
  in the context of rule-based constraint programming. We
  systematically derive a scheduler for them from a generic iteration
  algorithm of Apt \cite{Apt00a}.  We apply this study to so-called
  membership rules of Apt and Monfroy \cite{AM01}.  This leads to an
  implementation that yields for these rules a considerably better
  performance than   their execution as standard \texttt{CHR} rules.
\end{abstract}


\section{Introduction}
\label{sec:introduction}

In this paper we are concerned with schedulers for a class of rules
that naturally arise in the context of constraint programming
represented by means of rule-based programming.  An example
of such rules are so-called \emph{membership rules}, introduced in Apt
and Monfroy \cite{AM01}.
Their relevance stems from the following observations there made for
constraint satisfaction problems (CSP's) with finite domains:
\begin{itemize}

\item constraint propagation can be naturally achieved by repeated
  application of the membership rules;

\item in particular the notion of hyper-arc consistency can be characterized in
  terms of the membership rules;

\item for constraints explicitly defined on small finite domains all
  valid membership rules can be automatically generated (For a most
  recent reference on the subject of such an automatic rule generation
  see Abdennadher and Rigotti \cite{AR01}.);

\item many rules of the \texttt{CHR} language (Constraint Handling
  Rules) of Fr\"{u}hwirth \cite{FruehwirthJLP98} that are used
  in specific constraint solvers are in fact membership rules.  Now,
  in the logic programming approach to constraint programming
  \texttt{CHR} is the language of choice to write constraint solvers.
\end{itemize}

In the resulting approach to constraint programming the computation
process is limited to a repeated application of the rules intertwined
with splitting (labeling).
So the viability of this approach crucially depends on the
availability of efficient schedulers for such rules.  This motivates
the work here reported.  We provide an abstract framework for such
schedulers and use it as a basis for an implementation.

The abstract framework is based on an appropriate modification of the
generic approach to constraint propagation algorithms introduced in
Apt \cite{Apt99b} and Apt \cite{Apt00a}.  In this framework one
proceeds in two steps.  First, a generic iteration algorithm on
partial orderings is introduced and proved correct in an abstract
setting. Then it is instantiated with specific partial orderings and
functions to obtain specific constraint propagation algorithms.
In this paper, as in Apt \cite{Apt00a},
we take into account information
about the scheduled functions.  Here we consider functions in the form
of the rules $b \ra g$, where $b$ and $g$ satisfy a number of natural
conditions.  We call such functions \emph{good rules}.  The relevant
observation is that membership rules are good rules.  Then we propose
a specific scheduler in the form of an algorithm {\Ralgo},
appropriate for good rules.

The implementation is provided as an \eclipse{} program that accepts
as input a set of membership rules and constructs an \eclipse{}
program that is the instantiation of the {\Ralgo} algorithm for this
set of rules.  As membership rules can be naturally represented as
\texttt{CHR} propagation rules, one can assess this implementation by
comparing it with the performance of the standard implementation of
membership rules in the \texttt{CHR} language. We found by means of
various benchmarks that our implementation is considerably faster than
\texttt{CHR}.

\texttt{CHR} is available in a number of languages including the
\eclipse{} and the Sicstus Prolog systems.
In both cases \texttt{CHR} programs are compiled into the source language.
There is also a recent implementation in Java, see \cite{AKSS01}.
A great deal of effort was spent on implementing \texttt{CHR}
efficiently. For an account of the most recent implementation see
Holzbaur et al.  \cite{Holzbaur:2001:OCC}.  Since, as already
mentioned above, many \texttt{CHR} rules are membership rules,
our approach provides a better implementation of a subset of \texttt{CHR}.
While being stricly smaller than full \texttt{CHR},
the actual class of relevant rules is wider than the class of membership rules.
The essential properties, such as monotonicity of condition and conclusion,
are enjoyed by many rules that describe constraint propagation.
This, hopefully, may lead to new insights into design and
implementation of languages appropriate for writing constraint
solvers.

It is important to stress that the discussed implementation was obtained by
starting from ``first principles'' in the form of a generic iteration
algorithm on an arbitrary partial ordering.  This shows the practical
benefits of studying the constraint propagation process on an abstract
level.

\section{Revisions of the Generic Iteration Algorithm}
\label{sec:revised}

\subsection{The Original Algorithm}

Let us begin our presentation with recalling the generic algorithm of
Apt \cite{Apt00a}. We slightly adjust the presentation to our purposes
by assuming that the considered partial ordering also has the
greatest element $\top$.

So we consider a partial ordering $(D, \po )$ with the
least element $\bot$ and the greatest element $\top$,
and a set of functions $F := \C{f_1, \LL , f_k}$
on $D$.  We are interested in functions that satisfy
the following two properties.
\begin{definition}
\mbox{}\\[-5mm]

\begin{itemize}
\item $f$ is called {\em inflationary\/}
if $x \po f(x)$ for all $x$.

\item $f$ is called {\em monotonic\/}
if $x \po y$ implies
$f(x) \po f(y)$ \\for all $x, y$.
\HB
\end{itemize}
\end{definition}

Then the following algorithm is used to compute the least common
fixpoint of the functions from $F$.

\begin{figure}[ht]
\begin{tabbing}
\= $d := \bot$; \\
\> $G := F$; \\ 
\> {\bf while} $G \neq \ES$ {\bf and} $d \neq \top$ {\bf do} \\
\> \quad choose $g \in G$; \\
\> \quad $G := G - \C{g}$; \\
\> \quad $G := G \cup \update(G,g,d)$; \\
\> \quad $d := g(d)$ \\
\> {\bf end}
\end{tabbing}
\vspace{\figalgogap}
\caption{Generic Iteration Algorithm ({\GIalgo})}
\label{gialgo}
\end{figure}

where for all $G,g,d$ the set of functions 
$\update(G,g,d)$ from $F$ is such that
\begin{description}
\item[{\bf A}] $\C{f \in F - G  \mid f(d) = d \A f(g(d)) \neq g(d)} \sse
 \update(G,g,d)$,

\item[{\bf B}] $g(d) = d$ implies that $\update(G,g,d) = \ES$,

\item[{\bf C}] $g(g(d)) \neq g(d)$ implies that $g \in \update(G,g,d)$.
\end{description}

Intuitively, assumption {\bf A} states that $\update(G,g,d)$ 
contains at least all the functions from $F - G$ for which the ``old value'',
$d$, is a fixpoint but the ``new value'', $g(d)$, is not.  So at each
loop iteration such functions are added to the set $G$.  In turn,
assumption {\bf B} states that no functions are added to $G$ in case
the value of $d$ did not change.  Assumption {\bf C} provides
information when $g$ is to be added back to $G$ as this information is
not provided by {\bf A}.
On the whole, the idea is to keep in $G$ at least all functions $f$
for which the current value of $d$ is not a fixpoint.

The use of the condition $d \neq \top$, absent in the original
presentation, allows us to leave the \textbf{while} loop earlier.
Our interest in the {\GIalgo} algorithm is clarified by the
following result.

\begin{theorem}[Correctness] \label{thm:GI}
  Suppose that all functions in $F$ are inflationary and monotonic and
  that $(D, \po)$ is finite and has the least element $\bot$ and the
  greatest element $\top$.  Then every execution of the {\GIalgo}
  algorithm terminates and computes in $d$ the least common fixpoint
  of the functions from $F$.
\end{theorem}
\begin{proof}
(Sketch). The following statement is an invariant of the \textbf{while}
loop of the algorithm:
\[
(\fa f \in F - G \  f(d) = d) \A (\fa f \in F \ f(\top) = \top).
\]
This implies that the algorithm computes in $d$ a common fixpoint
of the functions from $F$. The fact that this is the least common
fixpoint follows from the assumption that all functions are monotonic.

In turn, termination is established by considering
the lexicographic ordering of the strict partial orderings
$(D, \sqsupset)$ and $({\cal N}, <)$,
defined on the elements of $D \times {\cal N}$ by
\[
(d_1, n_1) <_{lex} (d_2, n_2)\ {\rm iff} \ d_1 \sqsupset d_2
        \ {\rm or}\ ( d_1 = d_2 \ {\rm and}\ n_1 < n_2).
\]
Then with each {\bf while} loop iteration of the algorithm the pair
$(d, card \: G)$, where $card \: G$ is the cardinality of the set $G$,
strictly decreases in the ordering $<_{lex}$.
\end{proof}

\subsection{Removing Functions}

We now revise the {\GIalgo} algorithm by modifying dynamically the set
of functions that are being scheduled. The idea is that, whenever
possible, we remove functions from the set $F$.  This will allow us to
exit the loop earlier which speeds up the execution of the algorithm.

To realize this idea we proceed as follows.
First, we introduce the following property that will be satisfied by
the considered functions.

\begin{definition}
Suppose $d \in D$ and $f \in F$.
We say that $f$ is {\em stable above $d$\/} 
if $d \po e$ implies $f(e) = e$. We then say that $f$ is {\em stable}
if it is stable above $f(d)$, for all $d$.
\HB
\end{definition}

That is, $f$ is stable if for all $d$ and $e$, $f(d) \po e$ implies
$f(e) = e$.  So stability implies idempotence, which means that
$f(f(d)) = f(d)$, for all $d$.
Moreover, if $d$ and $f(d)$ are comparable
for all $d$, then stability implies inflationarity.
Indeed, if d \po $f(d)$, then the claim holds vacuously.
And if $f(d) \po d$, then by stability $f(d) = d$.

Next, we assume that for each function $g \in F$ and each element $d
\in D$, two lists of functions from $F$ are given, $\friends(g,d)$ and
$\obviated(g,d)$ that satisfy the following condition
\begin{eqnarray}
&& \fa d \: \fa e \sqsupseteq g \circ g_1 \circ \LL \circ g_k(d)%
\nonumber\\&&\quad
\fa f \in \friends(g,d) \cup \obviated(g,d) \: (f(e) = e)
\hspace{3em}
\label{eq:stable}
\end{eqnarray}
where $\friends(g,d) = [g_1, \LL, g_k]$.

That is, for all $d$, each function $f$ in $\friends(g,d) \cup \obviated(g,d)$
 is stable
above $g \circ g_1 \circ \LL \circ g_k(d)$.

Now, we modify the {\GIalgo}
algorithm in such a way that each application of $g$ to $d$ will be
immediately followed by the applications of all functions from
$\friends(g,d)$ and by a removal of the functions from $\friends(g,d)$
and from $\obviated(g,d)$ both from $F$ and $G$.
This modified algorithm is shown in Fig.~\ref{rgialgo}.
To keep the notation uniform we identified at some places the
lists $\friends(g,d)$ and $\obviated(g,d)$ with the sets.

\begin{figure}[ht]
\begin{tabbing}
\= $d := \bot$; \\
\> $F_0 := F$; \\
\> $G := F$; \\
\> {\bf while} $G \neq \ES$ {\bf and} $d \neq \top$ {\bf do} \\
\> \quad choose $g \in G$; \\
\> \quad $G := G - \C{g}$; \\
\> \quad $F := F - (\friends(g,d) \cup \obviated(g,d))$; \\
\> \quad $G := G - (\friends(g,d) \cup \obviated(g,d))$; \\
\> \quad $G$ \= $:= G \cup \update(G,h,d)$, \\
\>           \> where $h =  g \circ g_1 \circ \LL \circ g_k$ and $\friends(g,d)
 = [g_1, \LL, g_k]$;  \\
\> \quad $d :=h(d)$ \\
\> {\bf end}
\end{tabbing}
\vspace{\figalgogap}
\caption{Revised Generic Iteration Algorithm ({\RGIalgo})}
\label{rgialgo}
\end{figure}

The following result then shows correctness of this algorithm.

\begin{theorem}
  Suppose that all functions in $F$ are inflationary and monotonic and
  that $(D, \po)$ is finite and has the least element $\bot$ and the
  greatest element $\top$.  Additionally, suppose that for each
  function $g \in F$ and $d \in D$ two lists of functions from $F$ are
  given, $\friends(g,d)$ and $\obviated(g,d)$ such that condition
  (\ref{eq:stable}) holds.

Then the Correctness Theorem \ref{thm:GI} holds with the {\GIalgo} algorithm
replaced by the {\RGIalgo} algorithm.
\end{theorem}
\begin{proof}
In view of condition (\ref{eq:stable}) the following statement is an
invariant of the \textbf{while} loop:
\begin{eqnarray}
&\fa f \in F - G \: (f(d) = d) \;\A\; \fa f \in F \ (f(\top) = \top) \;\A%
\nonumber\\
&\fa f \in F_0 - F \ \fa e \sqsupseteq d \ (f(e) = e).
  \label{eq:del}
\end{eqnarray}
So upon termination of the algorithm the conjunction of this invariant with
the negation of the loop condition, i.e.,
\[
G = \ES \;\Or\; d = \top
\]
holds, which implies that $\fa f \in F_0 \: (f(d) = d)$.

The rest of the proof is the same.
\end{proof}

In the next section we shall focus on functions that are in a special
form. For these functions we shall show how to construct specific lists
$\friends(g,d)$ and $\obviated(g,d)$.

\subsection{Functions in the Form of Rules}

In what follows we consider the situation when the scheduled functions
are of a specific form $b \ra g$, where $b$ is a \emph{condition} and
$g$ a function, that we call a \emph{body}. We call such functions
\emph{rules}.

First, we explain how rules are applied.  Given an element $d$ of $D$,
a condition $b$ evaluates in $d$ to either true or false,
denoted $\holds(b,d)$ and $\nholds(b,d)$, resp.

Given a rule $b \ra g$ we define then its application as follows:
\[
(b \ra g)(d) :=
\left\{
\begin{array}{ll}
 g(d) & \mbox{if } \holds(b,d) \\
 d    & \mbox{if } \nholds(b,d) \enspace.
\end{array}
\right.
\]

The rules introduced in the next section will be of a specific type.

\begin{definition}
Consider a partial ordering $(D, \po )$.
\begin{itemize}

\item We say that a condition $b$ is \emph{monotonic} if
$\holds(b,d)$ and $d \po e$ implies $\holds(b,e)$,
for all $d, e$.

\item We say that a condition $b$ is \emph{precise} if
the least $d$ exists such that $\holds(b,d)$.
We call then $d$ the \emph{witness} for $b$.

\item  We call a rule $b \ra g$ \emph{good} if $b$ is monotonic and precise
and $g$ is stable.
\HB
\end{itemize}
\end{definition}

When all rules are good, we can modify the {\RGIalgo} algorithm by
taking into account that an application of a rule is a two step
process: testing of the condition followed by a conditional
application of the body. This will allow us to construct the lists
$\friends(g,d)$ and $\obviated(g,d)$ before the execution of the
algorithm, without using the parameter $d$. Moreover, the list
$\friends(g)$ can be constructed in such a way that the conditions of
its rules do not need to evaluated at the moment they are applied, as
they will all hold.  The details of a specific construction that we
shall use here will be given in a moment, once we identify the
condition that is crucial for the correctness.
This revision of the {\RGIalgo} algorithm is given in
Fig.~\ref{ralgo}.

\begin{figure}[ht]
\begin{tabbing}
\= $d := \bot$; \\
\> $F_0 := F$; \\
\> $G := F$; \\
\> {\bf while} $G \neq \ES$ {\bf and} $d \neq \top$ {\bf do} \\
\> \quad choose $f \in G$; suppose $f$ is $b \ra g$; \\
\> \quad $G := G - \C{b \ra g}$; \\
\> \quad {\bf if}  $\holds(b, d)$ {\bf then} \\
\> \quad \quad $F := F - (\friends(b \ra g) \cup \obviated(b \ra g))$; \\
\> \quad \quad $G := G - (\friends(b \ra g) \cup \obviated(b \ra g))$; \\
\> \quad \quad $G$ \= $:= G \cup \update(G,h,d)$, \\
\>                 \> where $h = g \circ g_1 \circ \LL \circ g_k$\\
\>                 \> and $\friends(b \ra g) = [b_1 \ra g_1, \LL, b_k \ra g_k]$;
  \\
\> \quad \quad $d :=h(d)$ \\
\> \quad {\bf else} \\
\> \quad \quad {\bf if}  $\forall e \sqsupseteq d \; \nholds(b, e)$ {\bf then}\\
\> \quad \quad \quad $F := F - \{ b \ra g \}$ \\
\> \quad \quad {\bf end} \\
\> \quad {\bf end} \\
\> {\bf end}
\end{tabbing}
\vspace{\figalgogap}
\caption{Rules Algorithm ({\Ralgo})}
\label{ralgo}
\end{figure}

Again, we are interested in identifying conditions under which the
Correctness Theorem \ref{thm:GI} holds with the {\GIalgo} algorithm
replaced by the {\Ralgo} algorithm.  To this end, given a rule $b \ra g$
in $F$ and $d \in D$, define as follows:

$
\begin{array}{l}
\friends(b \ra g, d) :=
\left\{%
\begin{array}{ll}
\friends(b \ra g) & \mbox{if } \holds(b, d)\\{}
[ \ ]    & \mbox{if } \nholds(b, d)
\end{array}%
\right.
\\\\
\obviated(b \ra g, d) :=
\left\{%
\begin{array}{ll}
\obviated(b \ra g) & \mbox{if } \holds(b, d)\\{}
[ b \ra g ]        & \parbox{30mm}{if  $\forall e \sqsupseteq d$\\\qquad\
 $\nholds(b, e)$}\\{}
[ \ ]     & \mbox{otherwise}
\end{array}%
\right.
\end{array}
$

We now have the following counterpart of the Correctness Theorem~\ref{thm:GI}.
\begin{theorem}[Correctness] \label{thm:R}
  Suppose that all functions in $F$ are good rules of the form $b \ra
  g$, where $g$ is inflationary and monotonic, and that $(D, \po)$ is
  finite and has the least element $\bot$ and the greatest element
  $\top$.  Further, assume that for each rule $b \ra g$ the lists
  $\friends(b \ra g,d)$ and $\obviated(b \ra g,d)$ defined as above
  satisfy condition~(\ref{eq:stable}) and the following condition:
\begin{eqnarray}
&\fa d (b' \ra g' \in \friends(b \ra g) \quad\A\quad \\%
&\holds(b, d) \ra \fa e \sqsupseteq g(d) \ \holds(b', e) \enspace.
\nonumber
\label{eq:friends}
\end{eqnarray}
Then the Correctness Theorem \ref{thm:GI} holds with the {\GIalgo}
algorithm replaced by the {\Ralgo}~algorithm.
\end{theorem}
\begin{proof}
It suffices to show that the {\Ralgo} algorithm is an instance of the
{\RGIalgo} algorithm.
On the account of condition (\ref{eq:friends}) and the fact that the
rule bodies are inflationary functions, $\holds(b, d)$
implies that
\[
((b \ra g) \circ (b_1 \ra g_1) \circ \LL \circ (b_k \ra g_k))(d) = (g \circ g_1
 \circ \LL \circ g_k)(d),
\]
where $\friends(b \ra g) = [b_1 \ra g_1, \LL, b_k \ra g_k]$.
This takes care of the situation when if $\holds(b, d)$.

In turn, the definition of $\friends(b \ra g,d)$ and $\obviated(b \ra g,d)$
and assumption \textbf{B} take care of the situation when if $\nholds(b, d)$.
When the condition $b$ fails for all $e \sqsupseteq d$, then we can
conclude that for all such $e$ we have $(b \ra g)(e)=e$.  This allows
us to remove at that point of the execution the rule $b \ra g$ from
the set $F$.  This amounts to adding $b \ra g$ to the set
$\obviated(b \ra g, d)$ at runtime. Note that
condition~(\ref{eq:stable}) is then satisfied.
\end{proof}

We now provide an explicit construction of the lists $\friends$ and
$\obviated$ for a rule $b \ra g$ in the form of the
algorithm in Fig.~\ref{foalgo}.  $\mathtt{GI}(d)$ stands here for the {\GIalgo}
algorithm activated with $\bot$ replaced by $d$ and the considered set
of rules as the set of functions $F$.
Further, given an execution of $\mathtt{GI}(e)$,
we call here a rule $g$ \emph{relevant} if at some point $g(d) \neq d$ holds
after the ``choose $g \in G$'' action.

\begin{figure}[ht]
\begin{tabbing}
\= $e := \mbox{witness of } b$;\\
\> $e := \mathtt{GI}(g(e))$; \\
\> $\friends(b \ra g) :=$ \=list of the relevant rules $h \in F$\\\>\> in the
 execution of $\mathtt{GI}(g(e))$;\\
\> $\obviated(b \ra g) := [\ ]$; \\
\> {\bf for each} $(b' \ra g') \in F - \friends(b \ra g)$ {\bf do} \\
\> \quad {\bf if}  $g'(e) = e$ {\bf or}  $\fa e' \sqsupseteq e \ \nholds(b',
 e')$ {\bf then} \\
\> \quad \quad $\obviated(b \ra g) := [b' \ra g' | \obviated(b \ra g)]$ \\
\> \quad {\bf end} \\
\> {\bf end}
\end{tabbing}
\vspace{\figalgogap}
\caption{Friends and Obviated Algorithm ({\FOalgo})}
\label{foalgo}
\end{figure}

Note that $b \ra g \not\in \friends(b \ra g)$
since $b \ra g$ is a good rule,
while \mbox{$b \ra g \in \obviated(b \ra g)$} since
by the stability of~$g$  $g(e) = e$ holds.

The following observation now shows the adequacy of the {\FOalgo} algorithm
for our purposes.

\begin{lemma}
Upon termination of the {\FOalgo} algorithm
conditions (\ref{eq:stable}) and (\ref{eq:friends}) hold,
where the lists $\friends(b \ra g,d)$ and $\obviated(b \ra g,d)$
are defined as before Theorem \ref{thm:R}.
\HB
\end{lemma}

Let us summarize now the findings of this section that culminated in
the {\Ralgo} algorithm.  Assume that all functions are of the form of
the rules satisfying the conditions of the Correctness Theorem
\ref{thm:R}. Then in the {\Ralgo} algorithm, each time the evaluation
of the condition $b$ of the selected rule $b \ra g$  succeeds,

\begin{itemize}
\item the rules in the list $\friends(b \ra g)$ are applied
directly without testing the value of their conditions,

\item the rules in $\friends(b \ra g) \cup \obviated(b \ra g)$
are permanently removed from the current set of functions $G$
and from $F$.
\end{itemize}

\subsection{Recomputing of the Least Fixpoints}

Another important optimization takes place when the {\Ralgo} algorithm is
repeatedly applied to compute the least fixpoint.  More specifically,
consider the following sequence of actions:
\begin{itemize}
\item we compute the least common fixpoint $d$ of the functions from $F$,

\item we move from $d$ to an element $e$ such that $d \po e$,

\item  we compute the least common fixpoint above $e$ of
the functions from $F$.
\end{itemize}
Such a sequence of actions typically arises in the framework of CSP's,
further studied in Section
\ref{sec:concrete}.  The computation of the least common fixpoint $d$
of the functions from $F$ corresponds there to the constraint
propagation process for a constraint $C$.  The moving from $d$ to
$e$ such that $d \po e$ corresponds to splitting or constraint
propagation involving another constraint, and the computation of the
least common fixpoint above $e$ of the functions from $F$ corresponds
to another round of constraint propagation for $C$.

Suppose now that we computed the least common fixpoint $d$ of the
functions from $F$ using the {\RGIalgo} algorithm or its modification
{\Ralgo} for the rules. During its execution we permanently removed
some functions from the set $F$.  Then these
functions are not needed for computing the least common fixpoint above
$e$ of the functions from $F$.  The precise statement is provided in
the following simple, yet crucial, theorem.

\newcommand{\Ffin} {\ensuremath{F_{\mathit{fin}}}}
\begin{theorem} \label{thm:repeated}
  Suppose that all functions in $F$ are inflationary and monotonic and
  that $(D, \po)$ is finite. Suppose that the least common fixpoint
  $d_0$ of the functions from $F$ is computed by means of the
  {\RGIalgo} algorithm or the {\Ralgo} algorithm.  Let \Ffin\ be the
  final value of the variable $F$ upon termination of the {\RGIalgo}
  algorithm or of the {\Ralgo} algorithm.

  Suppose now that $d_0 \po e$. Then the least common fixpoint $e_0$ above $e$
  of the functions from $F$ coincides with the least common fixpoint
  above $e$ of the functions from \Ffin.
\end{theorem}
\begin{proof}
Take a common fixpoint $e_1$ of the functions from \Ffin\
such that $e \po e_1$.  It suffices to prove that $e_1$ is common
fixpoint of the functions from $F$.  So take $f \in F - \Ffin$.
Since condition (\ref{eq:del}) is an invariant of the main
\textbf{while} loop of the {\RGIalgo} algorithm and of the {\Ralgo}
algorithm, it holds upon termination and consequently $f$ is stable
above $d_0$.  But $d_0 \po e$ and $e \po e_1$, so we conclude that
$f(e_1) = e_1$.
\end{proof}

Intuitively, this result means that if after splitting we relaunch the
same constraint propagation process we can disregard the removed
functions.

In the next section we instantiate the {\Ralgo} algorithm by a set
of rules that naturally arise in the context of constraint
satisfaction problems with finite domains. In Section
\ref{sec:implementation} we assess the practical impact of the
discussed optimizations.
\section{Concrete Framework}
\label{sec:concrete}

We now proceed with the main topic of this paper, the schedulers for
the rules that naturally arise in the context of constraint
satisfaction problems.  First we recall briefly the necessary
background on the constraint satisfaction problems.

\subsection{Constraint Satisfaction Problems}

Consider a sequence of variables $X := x_1, \LL, x_n$
where $n \geq 0$, with respective domains $D_1, \LL, D_n$
associated with them.  So each variable $x_i$ ranges over the domain
$D_i$.  By a {\em constraint} $C$ on $X$ we mean a subset of $D_1
\times \LL \times D_n$.
Given an element $d := d_1, \LL, d_n$ of $D_1 \times \LL \times D_n$
and a subsequence $Y := x_{i_1}, \LL, x_{i_\ell}$ of $X$ we denote by
$d[Y]$ the sequence $d_{i_1}, \LL, d_{i_{\ell}}$. In particular, for a
variable $x_i$ from $X$, $d[x_i]$ denotes $d_i$.

Recall that a {\em constraint satisfaction problem}, in short CSP, consists of
a finite sequence of variables $X$ with respective domains ${\cal
  D}$, together with a finite set $\cal C$ of constraints, each on a
subsequence of $X$. We write it as
\mbox{$\p{{\cal C}}{x_1 \in D_1, \LL, x_n \in D_n}$},
where $X := x_1, \LL, x_n$ and ${\cal D} :=
D_1, \LL, D_n$.

By a {\em solution\/} to $\p{{\cal C}}{x_1 \in D_1, \LL, x_n \in D_n}$
we mean an element $d \in D_1 \times \LL \times D_n$ such that for
each constraint $C \in {\cal C}$ on a sequence of variables $X$ we
have $d[X] \in C$.  We call a CSP {\em consistent\/} if it has a
solution.  Two CSP's with the same sequence of variables are called
{\em equivalent\/} if they have the same set of solutions.

\subsection{Partial Orderings}

With each CSP ${\cal P} := \p{{\cal C}}{x_1 \in D_1, \LL, x_n \in
  D_n}$ we associate now a specific partial ordering.  Initially we
take the Cartesian product of the partial orderings
\mbox{$({\cal P}(D_1), \supseteq ), \LL, ({\cal P}(D_n), \supseteq)$}.
So this ordering is of the form
\[
({\cal P}(D_1) \times  \LL \times {\cal P}(D_n), \supseteq)
\]
where we interpret $\supseteq$ as the 
the Cartesian product of the reversed subset ordering.
The elements of this partial ordering are sequences
$(E_1, \LL, E_n)$ of respective subsets of $(D_1, \LL, D_n)$ ordered
by the componentwise reversed subset ordering. Note that in this
ordering $(D_1, \LL, D_n)$ is the least element while
\[
\underbrace{(\ES, \LL, \ES)}_{\mbox{$n$ times}}
\]
is the greatest element. However, we would like to identify with the
greatest element all sequences that contain as an element the empty
set.  So we divide the above partial ordering by the equivalence
relation $R_{\ES}$ according to which
\begin{eqnarray*}
&(E_1, \LL, E_n)  \ R_{\ES} \ (F_1, \LL, F_n) \qquad\mbox{iff}\\
&(E_1, \LL, E_n) = (F_1, \LL, F_n)
\mbox{ or } (\te i \: E_i = \ES
\mbox{ and } \te j \: F_j = \ES).
\end{eqnarray*}
It is straightforward to see that $R_{\ES}$ is indeed an equivalence relation.

In the resulting quotient ordering there are two types of elements: the
sequences $(E_1, \LL, E_n)$ that do not contain the empty set as an
element, that we continue to present in the usual way with the
understanding that now each of the listed sets is non-empty, and one
``special'' element equal to the equivalence class consisting of all
sequences that contain the empty set as an element. This equivalence
class is the greatest element in the resulting ordering, so we denote
it by $\top$.  In what follows we denote this partial ordering by
$(D_{\cal P}, \po )$.

\subsection{Membership Rules}

Fix now a specific CSP ${\cal P} := \p{{\cal C}}{x_1 \in D_1, \LL, x_n \in D_n}$
with finite domains. We now recall the
rules introduced in Apt and Monfroy \cite{AM01}.%
\footnote
{In our presentation we slightly relax the original syntactic restrictions.}
They are called {\em membership rules} and are of the form
\[
y_1 \in S_1, \LL, y_k \in S_k \ra z_1 \neq a_1, \LL, z_m \neq a_m,
\]
where
\begin{itemize}
\item $y_1, \LL, y_k$ are pairwise different variables from the set $\C{x_1,
 \LL, x_n}$
  and $S_1, \LL, S_k$ are subsets of the respective variable domains,

\item  $z_1, \LL, z_m$ are variables from the set $\C{x_1, \LL, x_n}$
and $a_1, \LL, a_m$ are elements of the respective variable domains.
\end{itemize}
Note that we do not assume that the variables $z_1, \LL, z_m$
are pairwise different.

The computational interpretation of such a rule is:
\begin{quote}
if for $i \in [1..k]$ the current domain of the variable $y_i$
is included in the set $S_i$, then
for $j \in [1..m]$ remove the element $a_i$ from the
domain of $z_i$.
\end{quote}
When each set $S_i$ is a singleton, we call a membership rule an \emph{equality
 rule}.

Let us reformulate this interpretation so that it fits the framework
considered in the previous section.
To this end we need to clarify how to evaluate a condition, and how to
interpret a conclusion.
We start with the first item.
\begin{definition}
Given a variable $y$ with the domain $D_y$ and $E, S \supseteq D_y$ we define
\begin{tabbing}
\qquad $\holds(y \in S, E)$ \quad iff \quad $E \sse S$,
\end{tabbing}
and extend the definition to the elements of the considered ordering
$(D_{\cal P}, \po )$ by putting
\begin{tabbing}
\qquad $\holds(y \in S, (E_1, \LL, E_n))$ \quad iff \quad $E_k \sse S$,\\
\qquad\qquad\qquad where we assumed that $y$ is $x_k$,
\\\\
\qquad $\holds(y \in S, \top)$.
\end{tabbing}

Then we interpret a sequence $y_1 \in S_1, \LL, y_k \in S_k$ of conditions as a
 conjunction,
so by putting
\begin{tabbing}
\qquad $\holds(y_1 \in S_1, \LL, y_k \in S_k, (E_1, \LL, E_n))$
 \quad iff  \\
\qquad $\holds(y_i \in S_i, (E_1, \LL, E_n))$ for $i \in [1..k]$,
\\\\and\\\\
\qquad $\holds(y_1 \in S_1, \LL, y_k \in S_k, \top)$.
\qquad\qquad\quad\qquad\HB
\end{tabbing}
\end{definition}

Concerning the second item we proceed as follows.
\begin{definition}
  Given a variable $z$ with the domain $D_z$ we interpret the
  atomic formula $z \neq a$ as a function on ${\cal P}(D_z)$,
  defined by:
\[
(z \neq a)(E) := E - \C{a}.
\]
Then we extend this function to
the elements of the considered ordering $(D_{\cal P}, \po )$ as follows:
\begin{itemize}
\item on the elements of the form $(E_1, \LL, E_n)$
we use ``padding'', that
is we interpret it as the identity on the other components.
If the resulting sequence contains the empty set, we replace it by $\top$,
\item on the element $\top$ we put
$(z \neq a)(\top) := \top$
\end{itemize}
Finally, we interpret a sequence $z_1 \neq a_1, \LL, z_m \neq a_m$
of atomic formulas by interpreting each of them in turn.
\HB
\end{definition}

In view of the Correctness Theorem \ref{thm:R} the following
observation allows us to apply the {\Ralgo} algorithm when each
function is a membership rule and when for each rule
  $b \ra g$ the lists $\friends(b \ra g)$ and $\obviated(b \ra g)$ are
  constructed by the {\FOalgo} algorithm.

\begin{note}
Consider the partial ordering $(D_{\cal P}, \po )$.
\begin{enumerate}\smallromani
\item
Each membership rule is good.

\item
Each function $z_1 \neq a_1, \LL, z_m \neq a_m$ on $D_{\cal P}$ is
\begin{itemize}
\item inflationary,

\item monotonic.
\HB
\end{itemize}
\end{enumerate}
\end{note}

To be able to instantiate the algorithm {\Ralgo}
with the membership rules we still need to define the set
$\update(G,g,d)$. In our implementation we chose the following
simple definition:
\[
\update(G,b \ra g,d) :=
\left\{%
\begin{array}{ll}
F-G & \mbox{if } \holds(b, d) \A g(d) \neq d\\
\ES & \mbox{otherwise.}
\end{array}%
\right.
\]

To illustrate the intuition behind the use of the lists $\friends(b \ra g)$
and $\obviated(b \ra g)$ take the CSP~$\cal P :=$
\[
\p{{\cal C}}{x_1 \in \C{a,b,c}, x_2 \in \C{a,b,c}, x_3 \in \C{a,b,c}, x_4 \in
 \C{a,b,c}}
\]
and consider the membership rules
\begin{eqnarray*}
r_1 &:=& x_1 \in \C{a,b} \;\ra\; x_2 \neq a, x_4 \neq b,\\
r_2 &:=& x_1 \in \C{a,b}, x_2 \in \C{b,c} \;\ra\; x_3 \neq a,\\
r_3 &:=& x_2 \in \C{b} \;\ra\; x_3 \neq a, x_4 \neq b.
\end{eqnarray*}
Then upon application of rule $r_1$ rule $r_2$ can
be applied without evaluating its condition and subsequently
rule $r_3$ can be deleted without applying it.
So we can put rule $r_2$ into $\friends(r_1)$ and rule $r_3$ into
 $\obviated(r_1)$,
and this in fact is what the {\FOalgo} algorithm does.

\section{Implementation}
\label{sec:implementation}

In this section we discuss the implementation of the {\Ralgo}
algorithm for the membership rules and compare it by means of various
benchmarks with the \texttt{CHR} implementation in the \eclipse{} system.

\subsection{Modelling of the Membership Rules in \texttt{\large CHR}}

Following Apt and Monfroy \cite{AM01} the membership rules are
represented as \texttt{CHR} propagation rules with one head.  Recall
that the latter ones are of the form
\[
H ==> G_1, \ldots, G_l ~|~ B_1, \ldots, B_m.
\]
where
\begin{itemize}
\item $l \geq 0$, $m > 0$,

\item the atom $H$ of the \emph{head} refers to
the defined constraints,

\item the atoms of the \emph{guard} $G_1, \ldots,
G_l$ refer to Prolog relations or built-in constraints,

\item the atoms of the \emph{body} $B_1, \ldots, B_m$ are arbitrary atoms.
\end{itemize}

Further, recall that the \texttt{CHR} propagation rules with one head are
executed as follows.  First, given a query (that represents a CSP) the
variables of the rule are renamed to avoid variable clashes.  Then an
attempt is made to match the head of the rule against the first atom
of the query.  If it is successful and the guard of the instantiated version
of the rule succeeds, the instantiated version of the body of the rule is
executed. Otherwise the next rule is tried.

Finally, let us recall the representation of a membership rule as
a \texttt{CHR} propagation rule used in Apt and Monfroy \cite{AM01}.
Consider the membership rule
\[
y_1 \in S_1, \LL, y_k \in S_k \ra z_1 \neq a_1, \LL, z_m \neq a_m.
\]
related to the constraint \texttt{c} on the variables $X_1, \LL, X_n$.
We represent it as a \texttt{CHR} rule with the single head atom
$c(X_1, \LL, X_n)$ and guard atoms $\mathtt{in}(y_i, S_i)$
where the \texttt{in/2} predicate is defined by
\texttt{in(X,L) :- dom(X,D), subset(D,L)}.
The body consists of atomic calls $z_i\, \verb?##? \,a_i$.

In general, the application of a membership rule as defined in Section
\ref{sec:concrete} and the execution of its representation as a
\texttt{CHR} propagation rules coincide.  Moreover, by the semantics of
\texttt{CHR}, the \texttt{CHR} rules are repeatedly applied until a
fixpoint is reached.  So a repeated
application of a finite set of membership rules coincides with the
execution of the \texttt{CHR} program formed by the representations of
these membership rules as propagation rules.

\subsection{Benchmarks}

In our approach the repeated application of a finite set of
membership rules is realized by means of the {\Ralgo} algorithm of
Section \ref{sec:revised} implemented in \eclipse{}.
The compiler consists of about 1500 lines of code.
It accepts as input a set of membership rules, each represented as a
\texttt{CHR} propagation rule, and constructs an \eclipse{} program
that is the instantiation of the {\Ralgo} algorithm for this set of
rules.  As in \texttt{CHR}, for each constraint the set of rules that
refer to it is scheduled separately.

For each considered constraint we use rules
generated by a program discussed in \cite{AM01}.
Our compiler constructs then for each rule $g$ the lists
$\friends(g)$ and $\obviated(g)$ by executing the \FOalgo~algorithm
(essentially computing a fixpoint for each rule).
Time spent on this construction is comparable with rule generation time.

We chose benchmarks that embody several successive propagation steps,
i.\,e., propagation interleaved with domain splitting or labelling.
In Table \ref{tab:fixpoints} we list the results for selected single
constraints.  For each such constraint, say $C$ on a sequence of
variables $x_1, \LL, x_n$ with respective domains $D_1, \LL, D_n$, we
consider the CSP $\p{C}{x_1 \in D_1, \LL, x_n \in D_n}$ together with
randomized labelling. That is, the choices of a variable, value, and an
assignment or a removal of the value, are random.
The computation of only the solutions yields times that are insignificant,
so the test program computes also all intermediate fixpoints,
where some domains are not singleton sets.
Branching at these recorded points takes place only once, that is,
backtracking occurs immediately if a recorded point is encountered again.
In Table \ref{tab:atpg} we report the results for CSP's that formalize
sequential automatic test pattern generation for digital circuits
(ATPG).  These are rather large CSP's that employ the \texttt{and}
constraints of Table \ref{tab:fixpoints} and a number of other
constraints.  They are taken from a recent study by the first author
that will be reported elsewhere.

We measured the execution times for three rule schedulers:
the standard \texttt{CHR} representation of the rules,
the generic chaotic iteration algorithm {\GIalgo},
and its improved derivative~{\Ralgo}.
The codes of the latter two algorithms are both produced by
our compiler and are structurally equal, hence allow a direct
assessment of the improvements embodied in~{\Ralgo}.

An important point in the implementations is the question of when to
remove solved constraints from the constraint store.  The standard
\texttt{CHR} representation of membership rules
does so by containing, beside the propagation
rules, one \texttt{CHR} simplification rule for each tuple in
the constraint definition.  Once its variables are
assigned values that correspond to a tuple, the constraint is solved,
and removed from the store by the corresponding simplification rule.
This `solved' test takes place interleaved with propagation.
The implementations of {\GIalgo} and {\Ralgo}
check after closure under the propagation rules.  The constraint is
considered solved if all its variables are fixed, or, in the case of
{\Ralgo}, if the set $F$ of remaining rules is empty.

In the tables we provide for each constraint or CSP the ratio of
the execution times in seconds between,
first, {\Ralgo} and {\GIalgo}, and second,
{\Ralgo} and \texttt{CHR}.  This is followed by the
absolute times for {\Ralgo} and {\GIalgo} / \texttt{CHR}.

\newcommand{\xs}[1]     {{\small #1}}
\newcommand{\xp}[2]     {{\small {#1}\%/{#2}\%}}

\begin{table}[ht]
\begin{tabular*}{\columnwidth}{@{\extracolsep{\fill}}l@{}c@{}c@{}c@{}c@{}c@{}}%
\hline\hline\\[-3mm]
{\bfseries Const.}
        & \texttt{rcc8}
        & \texttt{fork}
        & \texttt{and3}
        & \texttt{and9}
        & \texttt{and11}
        \\\hline
\textsc{mem}\\
\  rel.
        & \xp{26}{11}
        & \xp{43}{40}
        & \xp{58}{47}
        & \xp{13}{6}
        & \xp{13}{3}
        \\
\  abs.
        & \xs{109}
        & \xs{0.23}
        & \xs{0.22}
        & \xs{70}
        & \xs{55.6}
        \\
        & \xs{419/950}
        & \xs{0.54/0.58}
        & \xs{0.38/0.47}
        & \xs{534/1096}
        & \xs{427/2077}
\\
\textsc{equ}\\
\  rel.
        & \xp{95}{100}
        & \xp{95}{89}
        & \xp{82}{74}
        & \xp{94}{97}
        & \xp{89}{94}
        \\[1mm]
\  abs.
        & \xs{323}
        & \xs{18.9}
        & \xs{0.31}
        & \xs{286}
        & \xs{299}
        \\
        & \xs{341/324}
        & \xs{19.9/21.2}
        & \xs{0.38/0.42}
        & \xs{303/294}
        & \xs{335/318}
\\\hline\hline\\
\end{tabular*}
\vspace{\figalgogap}
    \caption{Randomized search trees for constraints} 
    \label{tab:fixpoints}
  \centering
\end{table}

\newcommand{\xgg} {\hspace{1mm}}

\begin{table}[ht]
\begin{tabular*}{\columnwidth}
{@{\extracolsep{\fill}}l@{}c@{}c@{}c@{\extracolsep{0mm}}c}
\hline\hline\\[-3mm]
{\bfseries Logic}
        & 3-valued
        & 9-valued
        & 11-valued
7        \\[1mm]\hline
\textsc{mem}\\
\  relative
        & 64\% / 35\%
        & 71\% / 24\%
        & 85\% / 86\%
        \\
\  absolute
        & \xs{1.39 \xgg 2.16/4.01}
        & \xs{124  \xgg 175/509}
        & \xs{797  \xgg 933/3120}
        \\
\textsc{equ}\\
\  relative
        & 63\% / 70\%
        & 44\% / 59\%
        & 39\% / 48\%
        \\ 
\  absolute
        & \xs{0.72 \xgg 1.15/2.58}
        & \xs{2.40 \xgg 5.50/4.09}
        & \xs{12.3 \xgg 31.6/25.7}
\\\hline\hline\\
\end{tabular*}
\vspace{\figalgogap}
\caption{CSP's formalizing sequential ATPG}
\label{tab:atpg}
\centering
\end{table}

\subsection{Recomputing of the Least Fixpoints}

Finally, let us illustrate the impact of the permanent removal of the
rules during the least fixpoint computation, achieved here by the use
of the lists $\friends(g)$ and $\obviated(g)$.  Given a set $F$ of
rules call a rule $g \in F$ \emph{solving} if $\friends(g) \cup
\obviated(g) = F$.

Take now as an example the equivalence relation $\equiv$ from three valued logic
 of
Kleene \cite{Kle52}[page 334] that consists of three values, t (true),
f (false) and u (unknown).
It is defined by the truth table
\[
\begin{array}{|c|ccc|} \hline
\equiv & $t$ & $f$ & $u$ \\ \hline
$t$ & $t$ & $f$ & $u$ \\
$f$ & $f$ & $t$ & $u$ \\
$u$ & $u$ & $u$ & $u$ \\
\hline
\end{array}
\]
The program of Apt and Monfroy \cite{AM01} generates for it 26 minimal
valid membership rules. Out of them 12 are solving rules. For the
remaining rules the sizes of the set $\mathit{friends} \cup
\mathit{obviated}$ are: 17 (for 8 rules), 14 (for 4 rules), and 6 (for
2 rules).

In the {\Ralgo} algorithm a selection of a solving rule leads
directly to the termination ($G = \ES$) and to a reduction of the set
$F$ to $\ES$.  For other rules also a considerable simplification in the
computation takes place. For example, one of the 8 rules with 17 rules in
its set $\mathit{friends} \cup \mathit{obviated}$ is
\[
r:= x \in \{0\}, z \in \{0, u\} \ra y \not= 0.
\]
Consider the CSP
$\p{\equiv}{x \in \{0\}, y \in \{0,1,u\}, z \in \{0,u\}}$.
In the {\Ralgo} algorithm the selection
of $r$ is followed by the application of the rules in
$\friends$ and the removal of the rules in $\friends \cup \obviated$.
This brings the number of the considered rules down to $26 - 17 =
9$. The {\Ralgo}~algorithm subsequently discovers that none of these
nine rules is applicable at this point, so this set $F$ remains upon
termination.  Then in a subsequent constraint propagation phase,
launched after splitting or after constraint propagation involving
another constraint, the fixpoint computation by means of the
{\Ralgo} algorithm involves only these nine rules instead of the
initial set of 26 rules. For solving rules, this fixpoint computation
immediately terminates.

Interestingly, as Table \ref{tab:solving} shows, the solving rules
occur quite frequently.  We list there for each constraint and each
type of rules the number of solving rules divided by the total number
of rules, followed in a new line by the average number of rules in the
set $\friends(g) \cup \obviated(g)$.

\begin{table}[hb]
\begin{tabular*}{\columnwidth}{@{\extracolsep{\fill}}l@{}c@{}c@{}c@{}c@{}c@{}c@{}c@{}}%
\hline\hline
                & \texttt{and2}
                & \texttt{and3}
                & \texttt{and9}
                & \texttt{and11}
                & \texttt{fork}
                & \texttt{rcc8}
                & \texttt{allen}
                \\\hline\\[-2mm]
\textsc{equ} 
                & 6/6
                & 13/16
                & 113/134
                & 129/153
                & 9/12
                & 183/183
                & 498/498
                \\
                & 6
                & 14
                & 130
                & 148
                & 11
                & 183
                & 498
                \\[0mm]
\textsc{mem} 
                & 6/6
                & 4/13
                & 72/1294
                & 196/4656
                & 0/24
                & 0/912
                & -/26446    
                \\
                & 6
                & 7
                & 810
                & 3156
                & 9
                & 556
                & -          
                \\[0mm]
\hline\hline\\
\end{tabular*}
\vspace{\figalgogap}
\caption{Solving rules}
    \label{tab:solving}
\end{table}

The \texttt{fork} constraint is taken from the Waltz language for the
analysis of polyhedral scenes. The \texttt{rcc8} is the composition
table for the Region Connection Calculus with 8 relations from
Egenhofer \cite{Ege91}. It is remarkable that all its 183 minimal
valid equality rules are solving.  While none of its 912 minimal valid
membership rule for \texttt{rcc8} is solving, on the average the set
$\friends(g) \cup \obviated(g)$ contains 556 membership
rules.  Also all 498 minimal valid equality rules for the
\texttt{allen} constraint, that represents the composition table for
Allen's qualitative temporal reasoning, are solving.
\balancecolumns
The number of
minimal valid membership rules exceeds 26,000 and consequently they
are too costly to analyze.


The savings obtained by means of the lists $\friends(g)$ and $\obviated(g)$
are orthogonal to the ones obtained by a transformation of the \texttt{CHR}
propagation rules into the simplification rules discussed in
Abdennadher and Rigotti \cite{AR01}.
We think that there is a relation between two approaches that we plan to
study closer.

\section*{Acknowledgments}

We thank Christian Holzbaur and Eric Monfroy for helpful discussions
on the implementation and on an early version of this paper,
and the referees for useful comments.

\bibliographystyle{plain}

\bibliography{/ufs/apt/book-ao-2nd/apt,/ufs/apt/esprit/esprit,
/ufs/apt/esprit/chapter3,/ufs/apt/bib/clp2,/ufs/apt/bib/clp1,
/ufs/apt/book-lp/man1,/ufs/apt/book-lp/man2,/ufs/apt/book-lp/man3,
/ufs/apt/book-lp/ref1,/ufs/apt/book-lp/ref2,/ufs/apt/bib/99}

\begin{thebibliography}{1}

\bibitem{AKSS01}
S.~Abdennadher, E.~{Kr\"{a}mer}, M.~Saft, and M.~Schmaus.
\newblock {JACK: A Java Constraint Kit}.
\newblock In {\em International Workshop on Functional and (Constraint) Logic
  Programming (WFLP 2001), Kiel, 2001}, 2001.

\bibitem{AR01}
S.~Abdennadher and C.~Rigotti.
\newblock Using confluence to generate rule-based constraint solvers.
\newblock In {\em Proceedings of the 3rd Int. Conf. on Principles and Practice
  of Declarative Programming (PPDP 2001), Firenze, Italy}, September 2001.

\bibitem{Apt99b}
K.~R. Apt.
\newblock The essence of constraint propagation.
\newblock {\em Theoretical Computer Science}, 221(1--2):179--210, 1999.
\newblock Available via \verb+http://arXiv.org/archive/cs/+.

\bibitem{Apt00a}
K.~R. Apt.
\newblock The role of commutativity in constraint propagation algorithms.
\newblock {\em ACM Transactions on Programming Languages and Systems},
  22(6):1002--1036, 2000.
\newblock Available via \verb+http://arXiv.org/archive/cs/+.

\bibitem{AM01}
K.~R. Apt and E.~Monfroy.
\newblock Constraint programming viewed as rule-based programming.
\newblock {\em Theory and Practice of Logic Programming}, 1(6):713--750, 2001.
\newblock Available via \verb+http://arXiv.org/archive/cs/+.

\bibitem{Ege91}
M.~Egenhofer.
\newblock Reasoning about binary topological relations.
\newblock In O.~G{\"{u}}nther and H.-J. Schek, editors, {\em Proceedings of the
  2nd International Symposium on Large Spatial Databases ({SSD})}, volume 525,
  pages 143--160. Springer-Verlag, 1991.

\bibitem{FruehwirthJLP98}
T.~Fr{\"{u}}hwirth.
\newblock Theory and practice of constraint handling rules.
\newblock {\em Journal of Logic Programming}, 37(1--3):95--138, October 1998.
\newblock Special Issue on Constraint Logic Programming (P. Stuckey and K.
  Marriot, Eds.).

\bibitem{Holzbaur:2001:OCC}
C.~Holzbaur, M.~Garc{\'\i}a de~la Banda, D.~Jeffery, and P.~J. Stuckey.
\newblock Optimizing compilation of constraint handling rules.
\newblock In {\em Proceedings of the 2001 International Conference on Logic
  Programming}, volume 2237 of {\em Lecture Notes in Computer Science}, pages
  74--89. Springer-Verlag, 2001.

\bibitem{Kle52}
S.~C. Kleene.
\newblock {\em Introduction to Metamathematics}.
\newblock van Nostrand, New York, 1952.

\end{thebibliography}

\end{document}